\documentclass[conference]{IEEEtran}

\IEEEoverridecommandlockouts

\usepackage{cite}

\usepackage{url}

\usepackage{booktabs}

\usepackage{array}

\usepackage{tabularx}

\usepackage{amsmath}

\usepackage{amssymb}

\usepackage{balance}
\usepackage{graphicx}
\usepackage{xcolor}

\newcommand{\seventhG}{7G}

\newcommand{\sixthG}{6G}

\newcommand{\fifthG}{5G}

\begin{document}

\title{Will there be a 7G?}

\author{
\IEEEauthorblockN{Adnan Aijaz}
\IEEEauthorblockA{Bristol Research and Innovation Laboratory, Toshiba Europe Ltd.,
Bristol, United Kingdom\\
Email: adnan.aijaz@toshiba-bril.com}
}

\maketitle

\begin{abstract}

The transition from 5G to 6G is becoming concrete: the ITU-R IMT-2030 framework has established the high-level vision and capability set for 6G, while 3GPP Release 21 has defined the path toward the first 6G specifications. This raises a deliberately provocative question for the research and standards communities: \emph{will there be a 7G}, and if so, what would justify it? This paper argues that 7G should not be treated as an inevitable numbering exercise or as a catalogue of more ambitious radio targets. Instead, its justification should depend on whether post-6G systems introduce needs or coordination problems that cannot be met by 6G/6G-Advanced, Wi-Fi, NTN, private cellular, neutral-host deployments, edge-cloud platforms, or complementary wireless and software-based systems. To support this assessment, the paper develops a readiness framework covering demand-led need, system-level discontinuity, coordination value, sustainability and circularity, trust, and geopolitical viability. It then applies the framework to candidate 7G discontinuities, including agentic network operation, RF-native computing, quantum-enabled interworking, policy-aware spectrum governance, grid-interactive infrastructure, outcome-assured services, and regionalized standards. The contribution is not a prediction of a fixed 7G architecture, but a structured basis for deciding whether 7G should become a distinct mobile generation, an extension of 6G evolution, or a broader post-6G infrastructure fabric.
\end{abstract}


\section{Introduction}

Mobile/Cellular communications generations have historically served two functions. First, they have introduced new technical capabilities: wider bandwidths, higher-order antenna systems, lower latency, new core-network architectures, and more efficient air interfaces. Second, and often less explicitly, they have coordinated investment cycles across standards bodies, regulators, spectrum agencies, semiconductor vendors, network equipment providers, mobile network operators (MNOs), device ecosystems, and application providers. In this sense, a ``G'' is not merely a radio interface; it is a coordination mechanism.

The question ``Will there be a 7G?'' is therefore not simply a speculative technology question. It is a question about whether the telecoms ecosystem will still benefit from bundling research, spectrum, equipment refreshes, device roadmaps, and regulatory decisions into a generational framework after 6G. This question is timely because 6G is no longer only a research label. The ITU-R IMT-2030 framework identifies overarching principles such as sustainability, security and resilience, connecting the unconnected, and ubiquitous intelligence~\cite{itu_m2160_2023,itu_imt2030_requirements_2026}. 3GPP has approved the Release 21 timeline for the first 6G specifications, with Stage-1 freeze in 2027, Stage-2 freeze in 2028, Stage-3 freeze in December 2028, and ASN.1/OpenAPI freeze in 2029~\cite{3gpp_rel21_2026}. The first 3GPP 6G RAN study on scenarios and requirements has also been approved~\cite{3gpp_tr38914_2026}.

At the same time, the 6G umbrella has become broad. It now spans AI-native network operation, integrated sensing and communications (ISAC), terrestrial and non-terrestrial integration, distributed cloud and edge computing, network automation, energy efficiency, and new security requirements~\cite{ngmn_6g_2023,kaushik_2023_isac}. If 6G already absorbs these themes, then 7G must answer a harder question: what remains sufficiently different to justify another generation? Higher data rates, lower latencies, and new high-frequency bands may be valuable, but not by themselves sufficient reasons for a new G-cycle. 

\subsection{Related Work}

A small but growing body of explicitly 7G-oriented literature has begun to appear. Early work has associated 7G with quantum-assisted optimization, quantum-secure networking, AI-enabled system-of-systems control, satellite integration, and mobile near-field terahertz communications~\cite{glisic_2024_7g_enablers,lorenzo_2025_quantum_network_design_6g7g,ahmed_2026_quantum_protocol_7g}. Related 6G surveys have also begun to speculate on what may follow 6G, including more pervasive intelligence, tighter integration of sensing and communication, and broader cyber-physical infrastructure. However, much of the emerging 7G literature remains technology-catalogue oriented: it identifies candidate enablers, but does not provide a systematic test for whether those enablers actually justify a new mobile generation.

\begin{figure}
\centering
\includegraphics[width=\columnwidth]{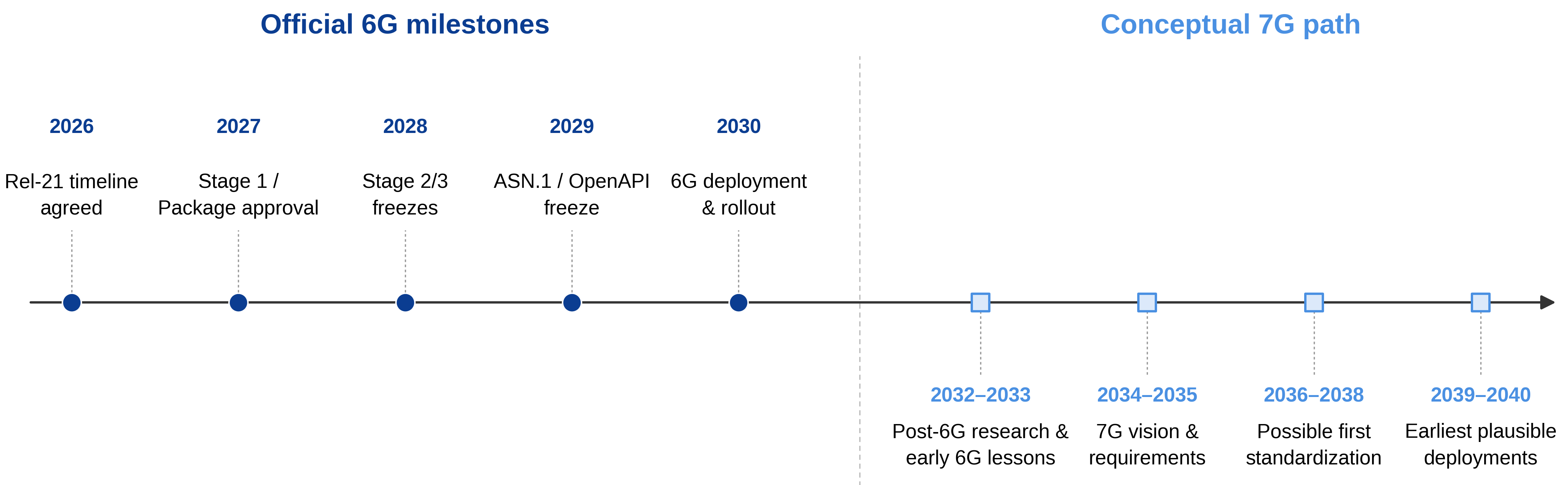}
\caption{Timeline for 7G; 6G milestones based on the 3GPP Release 21 timeline (agreed Jun'26); 7G milestones are illustrative. }
\label{timeline}
\vspace{-1.5em}
\end{figure}

%

\subsection{Contributions and Outline}

This paper is fundamentally different in scope and methodology. Rather than specifying a 7G air-interface, system architecture, or a technology stack, it asks whether 7G deserves to exist as a distinct standards generation at all.  The paper therefore treats 7G as a coordination problem across demand, system-level discontinuity, sustainability, trust, spectrum governance, business models, and geopolitics.

The paper makes three contributions. First, it distinguishes a 7G \emph{label} from 7G \emph{substance}, arguing that another standards cycle is not by itself a justification for another generation. Second, it proposes a readiness framework for testing whether post-6G capabilities require generational treatment. Third, it maps candidate 7G drivers to standards-facing gaps, including agentic operation, RF-native computing, quantum-enabled interworking, grid-interactive infrastructure, outcome assurance, sustainability, and regionalized standards.

The remainder of the paper is organized as follows. Section~II establishes the 6G baseline. Section~III introduces the readiness framework. Section~IV examines candidate sources of 7G differentiation. Sections~V--VII discuss business models, sustainability, and geopolitics. Section~VIII develops a 7G standardization agenda, and Section~IX concludes.

\section{The 6G Baseline for a 7G Debate}

A meaningful 7G discussion must begin by clarifying the 6G baseline. If 6G already includes a capability, then that capability alone cannot justify 7G unless it evolves into a fundamentally different operating model.

\subsection{IMT-2030 and 3GPP Timing}

Recommendation ITU-R M.2160 establishes the IMT-2030 framework for 6G and identifies usage scenarios and capabilities extending IMT-2020~\cite{itu_m2160_2023}. The published IMT-2030 capability set also gives useful quantitative context: target peak data rates are scenario-dependent at 50--200~Gbit/s, and spectrum efficiency is targeted at roughly 1.5--3 times IMT-2020~\cite{itu_imt2030_requirements_2026}. In public ITU material, 6G is framed around four overarching principles: sustainability, security and resilience, connecting the unconnected, and ubiquitous intelligence~\cite{itu_imt2030_requirements_2026}. This matters because sustainability and trust are no longer optional appendices; they are part of the formal design language.

3GPP Release 21 now gives 6G a concrete standardization horizon which paves the way for first-wave commercial 6G deployments around 2030.

\subsection{Technology Scope}

The 6G scope is broad and still unsettled. It includes the following families of capabilities:

\begin{itemize}

    \item \textbf{AI-native networking:} AI/ML is expected to support radio optimization, mobility, positioning, energy saving, network automation, and potentially air-interface functions~\cite{3gpp_ai_ml_nr}.

    \item \textbf{ISAC:} 6G is expected to combine communication and sensing, creating new system and RAN architecture questions~\cite{etsi_isac_2025}.

    \item \textbf{NTN integration:} Non-terrestrial networks already entered 3GPP Release 17 for NR NTN and IoT NTN, which means NTN is an evolution path from 5G to 6G rather than a purely future concept~\cite{3gpp_ntn_2024}.

    \item \textbf{Distributed cloud and edge:} 6G is expected to interact with distributed computing, AI inference, digital twins, and cyber-physical systems, although these ecosystems are not owned by cellular standards alone.

    \item \textbf{Quantum-safe security:} Post-quantum cryptography is already becoming deployable after NIST finalized the first three post-quantum standards in 2024~\cite{nist_pqc_2024}.

\end{itemize}

The breadth of this scope creates an important implication: 7G cannot be justified merely by listing more ambitious versions of 6G themes. It must demonstrate a new coordination problem or a new class of capability that cannot be handled through 6G-Advanced evolution.

\section{A 7G Readiness Framework}

This section proposes six tests for evaluating whether a post-6G capability should be treated as a candidate 7G driver. These tests are summarized in Table~\ref{tab:readiness}.

\begin{table*}[t]
\centering
\caption{A standards-centric readiness framework for assessing whether \seventhG{} is justified}
\label{tab:readiness}
\footnotesize
\setlength{\tabcolsep}{3pt}
\renewcommand{\arraystretch}{0.95}
\begin{tabularx}{\textwidth}{
>{\raggedright\arraybackslash}p{0.2\textwidth}
>{\raggedright\arraybackslash}X
>{\raggedright\arraybackslash}X}
\toprule
\textbf{Readiness test} & \textbf{Decision question} & \textbf{Implication for \seventhG{}} \\
\midrule
Demand-led need &
Who needs a new G, and what cannot be met by \sixthG{}/\sixthG{}-Advanced, Wi-Fi, NTN, private cellular, neutral hosts, or complementary wireless systems &
Anchor \seventhG{} in identifiable buyer and user needs, not supply-side R\&D cycles \\

System-level discontinuity &
Does the capability require new architecture, PHY assumptions, security, compute, spectrum, KPIs, or interfaces &
Keep incremental gains in \sixthG{} evolution; consider \seventhG{} only for true discontinuities \\

Coordination value &
Does a new G coordinate devices, spectrum, vendors, regulation, security, and investment better than continuous releases &
Use \seventhG{} only where standards create ecosystem scale, not new labels \\

Sustainability and circularity &
Can another generation be justified under carbon, energy, e-waste, and hardware-lifetime constraints &
Require absolute-impact metrics, circular procurement, lifecycle design, and grid-interactive operation \\

Trust and quantum-era security &
Does the network need a security model beyond classical and post-quantum upgrades &
Define crypto-agility, quantum-safe orchestration, and possible quantum-network interworking \\

Geopolitical viability &
Can a global standard survive divergent spectrum policy, AI governance, supply chains, and sovereignty requirements &
Expect global harmonization, regional profiles, or partial fragmentation \\
\bottomrule
\end{tabularx}
\end{table*}

\subsection{Demand-Led Need}

A post-6G generation should begin with demand, not technology. The relevant question is not whether researchers can imagine higher performance targets, but whether any stakeholder needs capabilities that cannot be addressed by 6G evolution. Candidate stakeholders include MNOs, enterprises, defence, public safety, utilities, neutral-host infrastructure providers, satellite operators, hyperscalers, industrial automation vendors, and governments.

The demand-led test is deliberately strict. A use-case should not count as a 7G driver if it can be better served by fibre, Wi-Fi, private networks, neutral-host infrastructure, satellite, or application-layer cloud services. This distinction is important because many emerging services associated with 6G such as robotics, distributed AI, digital twins, edge inference, and physical AI are evolving outside cellular standards. Mobile networks may support them, but need not define them.

\subsection{System-Level Discontinuity}

A new generation is more credible when it reflects a system-level discontinuity\footnote{A system-level discontinuity exists when a post-6G capability cannot be delivered through incremental enhancement of 6G/6G-Advanced, but instead requires new architectural assumptions, new KPIs, new interfaces, or new forms of cross-domain coordination.} rather than an accumulation of incremental enhancements.  Higher data rates, lower latency, better energy efficiency, or additional spectrum may be important, but they are not necessarily generational discontinuities. By contrast, radio components used as computing primitives, quantum-enabled network interworking, outcome-assured cyber-physical services, grid-interactive energy co-design, policy-aware spectrum governance, and deeply integrated sensing-computing-control loops may require standards coordination beyond conventional radio evolution.

\subsection{Coordination Value}

The strongest historical argument for a G-cycle is coordination. If 7G cannot coordinate investment, device roadmaps, spectrum policy, security baselines, testing, and interoperability better than a continuous release model, it may not be needed. This test is particularly important if networks become increasingly software-defined. A post-6G world may favor continuous cloud-style evolution rather than a large hardware-led generation.

\subsection{Sustainability and Circularity}

The sustainability test asks whether another generation can be environmentally justified. This is not limited to radio energy efficiency. Network energy, embodied carbon, site construction, device replacement, rare materials, and e-waste must all be considered. The operational scale is material: energy consumption is commonly reported as 20--40\% of mobile network OPEX, and public operator data cited by GSMA indicate that 73\% of network energy consumption is in the RAN, with core network, data centres, and operations accounting for the remainder~\cite{gsma_zain_energy_2023}. At the equipment level, Ericsson reports that base stations historically comprised around 80\% of RAN electricity use and that roughly 80\% of each base station's energy consumption was used in power amplifiers~\cite{ericsson_energy_2022}. Meanwhile, global e-waste reached 62 million tonnes in 2022, with only 22.3\% formally collected and recycled~\cite{itu_ewaste_2024}. If 7G accelerates hardware refreshes without lifecycle benefits, it may fail the sustainability test even if it improves bits per joule.

\subsection{Trust and Quantum-Era Security}

The quantum-era security test distinguishes between post-quantum secure 6G and potentially quantum-enabled 7G. NIST's final standards for ML-KEM, ML-DSA, and SLH-DSA provide a foundation for post-quantum cryptographic migration~\cite{nist_pqc_2024}. That migration is likely a 6G and critical-infrastructure task. A distinct 7G question is whether future networks need interworking with quantum key distribution, quantum random number generation, quantum repeaters, or quantum internet infrastructures~\cite{wehner_2018_quantum_internet}. This would move beyond securing classical networks against quantum attacks toward integrating classical and quantum network resources.

\subsection{Geopolitical Viability}

Previous generations benefited from global scale, even when regional spectrum and vendor preferences differed. The 7G question is whether that global model can survive increased divergence in spectrum policy, AI regulation, supply-chain security, sanctions, and national industrial policy. A fully fragmented 7G would damage device economies of scale and roaming. A more likely outcome may be a common technical baseline with regional security profiles, spectrum bands, AI governance constraints, and procurement rules. Such partial fragmentation still matters for standards.

\section{What could make 7G Different?}
This section examines candidate sources of 7G differentiation beyond 6G-Advanced. The aim is not to claim that all of them will define a future 7G system, but to identify where incremental 6G evolution may be insufficient. Each example highlights a possible shift in architecture, service model, control logic, trust assumptions, spectrum use, or infrastructure coordination that could require new standards treatment.

\subsection{From AI-Native to Agentic Network Operation}

%

\sixthG{} is already associated with AI-native networking, but this is not fully agentic operation. Near-term AI/ML work is bounded and function-specific: channel-state feedback, beam management, positioning, energy optimization, anomaly detection, and assurance~\cite{3gpp_ai_ml_nr}. These functions can be specified through measurable inputs, outputs, and conformance tests. Agentic systems involve goal interpretation, tool use, multi-step reasoning, API invocation, and cross-domain decisions, making them difficult to standardize in the first \sixthG{} cycle, where deterministic behaviour, bounded failures, repeatable testing, and accountability remain essential.

A possible \seventhG{} discontinuity is therefore not ``more AI'', but controlled agentic operation above real-time control loops. Open-ended agents are poorly suited to scheduling, handover, interference coordination, and ultra-low-latency control without strict safety envelopes. A plausible \seventhG{} architecture could use agents for planning, assurance, policy, energy optimization, spectrum negotiation, API orchestration, and service lifecycle management. The standards challenge is to define intent interfaces, authority boundaries, guardrails, auditability, rollback, explainability, and conformance tests for agent-assisted operation.

\subsection{Computing in Radio Components}

%
%

A more radical post-\sixthG{} idea is radio hardware that computes rather than only transports bits. Recent analog RF computing work uses RF front-end operations, including passive mixer multiplication, for energy-efficient edge AI over MU-MIMO systems~\cite{yu_2026_rf_computing}. This shifts the model from ``edge computing near the radio'' to ``computing in the radio.''

Such concepts are unlikely to enter first-wave \sixthG{} standards because they remain hardware-specific, model-dependent, and hard to express as interoperable radio procedures. They also blur the separation between PHY design, RF implementation, and application-layer AI. Open issues include calibration, analog noise, model distribution, privacy leakage, weight security, and coexistence.

If these concepts mature, \seventhG{} may need computation-centric PHY metrics such as inference accuracy, energy per inference, analog compute noise, and model freshness. The standards question is how to define waveforms, interfaces, calibration, and security profiles for radios that compute as well as communicate.

\subsection{From Quantum-Safe to Quantum-Enabled}

%
%

Post-quantum cryptography (PQC) can be introduced through crypto-agility and hybrid migration. Practical wireless deployment still brings larger handshakes, bandwidth overhead, and reliability challenges in constrained edge environments~\cite{aijaz_PQC}. This is an important \sixthG{} security upgrade, but not a generational discontinuity.

Quantum-enabled networking is different. It implies interworking among mobile, optical, satellite, quantum-key, and possibly quantum-repeater infrastructures. This is unlikely to enter first-wave \sixthG{} standards because deployment is sparse, interfaces are immature, and demand is concentrated in defence, government, finance, critical infrastructure, and high-value industrial networks.

The \seventhG{} question is therefore whether mobile systems remain classical networks protected by PQC, or expose quantum-secure connectivity as a service and interwork with sovereign quantum communication infrastructure. If so, standards would need quantum-key service exposure, hybrid trust anchors, cross-domain key management, assurance, and interoperability across mobile, optical, satellite, and quantum-network domains.

\subsection{Spectrum Governance Instead of More Spectrum}

Spectrum has always been central to mobile generations, but the post-6G spectrum story is unlikely to be a simple search for clean greenfield bands. WRC-23 expanded the mobile footprint in the upper 6~GHz range, with 6.425--7.125~GHz now harmonized across a footprint covering more than 80\% of the global population~\cite{gsma_upper6ghz_2025}. For 6G, the WRC-27 agenda includes studies related to IMT in bands such as 4.4--4.8~GHz, 7.125--8.4~GHz, and 14.8--15.35~GHz, with sharing and compatibility conditions under consideration~\cite{itu_wrc27_2026}. These bands involve incumbents and coexistence challenges, so nominal bandwidth alone is not a sufficient measure of spectrum readiness.


For \seventhG{}, the key discontinuity may not be a new clean band, but policy-aware access, dynamic sharing, regional fragmentation, multi-RAT aggregation, and coexistence with radar, fixed links, satellite, Wi-Fi, and private networks. The spectrum question therefore becomes governance-centric: how should networks reason about spectrum rights, interference externalities, and regional rules in real-time?

\subsection{From Energy-Efficient to Grid-Interactive Networks}

A further candidate discontinuity is the shift from energy-efficient to grid-interactive networking. Conventional green-network work reduces energy use through sleep modes, energy-aware scheduling, efficient power amplifiers, and traffic-load adaptation. Renewable-powered sites, battery-backed base stations, and telecom virtual power plants (VPPs) are gaining traction, but in the first \sixthG{} cycle they are likely to remain operator, infrastructure, or energy-market implementations rather than standardized radio-system features. They depend on local grid regulation, tariffs, site ownership, batteries, renewable availability, market rules, and operational risk models that sit largely outside cellular standardization.

Packetized energy management (PEM) has been proposed to represent flexible telecom demand as schedulable energy packets and aggregate PEM-enabled sites into telecoms VPPs~\cite{aijaz_2026_pem_6g}. Its significance for \seventhG{} is the broader architectural shift: base stations, batteries, renewables, and edge workloads become controllable energy assets rather than passive loads.

A \seventhG{} discontinuity would arise if this coordination moved from proprietary energy management into standardized network behaviour. Future systems may require interfaces between RAN control, edge-cloud orchestration, site energy systems, carbon-intensity signals, and grid-flexibility markets. If sustainability becomes a hard design constraint rather than a reporting metric, telco-energy co-design could become a non-radio driver for post-\sixthG{} standardization.

\subsection{A Computing-Sensing-Security Fabric}

ISAC, edge computing, distributed AI, and security are usually discussed as separate threads. A 7G-level discontinuity may occur if they become a single fabric. For example, a future network might jointly optimize radio sensing, local inference, digital twins, quantum-safe key distribution, and cyber-physical control. Such an architecture would be different from a network that simply transports data to cloud applications. The danger is overreach. Many of these ecosystems will evolve independently of mobile standards. A credible 7G standard should define interfaces and interworking functions rather than claiming ownership of robotics, AI platforms, cloud computing, or quantum infrastructure.

%
%
%
%
%
%
%
%
%

\section{Business Model and Stakeholder Implications}

The business case for \seventhG{} is uncertain because advanced \fifthG{} monetization remains uneven and early \sixthG{} commercial models are still forming. The lesson from \fifthG{} is that better connectivity does not automatically create better monetization. Enterprises buy productivity, reliability, safety, resilience, compliance, and efficiency, not radio features in isolation. The \sixthG{}--\seventhG{} transition may therefore be less about another connectivity tier and more about exposing, assuring, and monetizing operational outcomes.
This implies a shift from connectivity-centric to outcome-oriented service models. In industrial and robotic environments, value may be measured through task completion, deadline compliance, safety-zone assurance, time-to-service, mean time to assure, energy per task, and fleet productivity~\cite{aijaz_robotics_6g_2026}. Connectivity, positioning, sensing, edge computing, AI orchestration, security, and assurance may therefore need to be bundled into assured operational services.

For \seventhG{}, the business model may not be a conventional ``G'' model. If \seventhG{} becomes a wider infrastructure fabric, value may come from coordinating public mobile networks, private cellular, neutral hosts, NTN, edge-cloud platforms, sensing, quantum-safe security, and energy systems. Several models could emerge:

\begin{itemize}
    \item \textbf{Outcome-assured services:} charging for task completion, safety, deadline compliance, productivity, resilience, or mission success instead of data volume or generic connectivity.

    \item \textbf{Capability-as-a-service:} exposing positioning, sensing, compute proximity, trust, carbon profile, energy flexibility, and control-loop reliability through APIs.

    \item \textbf{Cyber-physical assurance platforms:} bundling network, edge, AI, sensing, and security for robotics, industrial automation, transport, healthcare, defence, and public safety.

    \item \textbf{Sovereign and resilient infrastructure services:} monetizing secure, regionally governed, quantum-safe, high-resilience connectivity for government, critical infrastructure, defence, finance, and emergency services.

    \item \textbf{Federated infrastructure brokerage:} coordinating MNO, neutral-host, private cellular, NTN, edge-cloud, fibre, sensing, and energy assets across heterogeneous infrastructure.

    \item \textbf{Sustainability and energy-flexibility services:} using carbon-aware orchestration, grid-interactive sites, storage, renewables, and telecom VPPs as service value.
\end{itemize}

\seventhG{}, if it emerges, may be justified less by consumer broadband than by standardizing how outcomes are requested, composed, assured, audited, and monetized. Applications may request robotic inspection, emergency corridors, quantum-safe industrial sessions, or carbon-constrained edge inference rather than slices. Such outcomes require coordination across radio, sensing, compute, security, energy, and governance.

The stakeholder set is broader than in MNO-centric generations: MNOs, neutral hosts, tower companies, private-network operators, satellite providers, cloud and edge providers, industrial integrators, robotics vendors, energy aggregators, security providers, regulators, and governments. Standardization must cover outcome KPIs, capability APIs, assurance telemetry, federation, liability, data governance, risk allocation, and settlement. The key \seventhG{} question is whether a generational framework is needed to coordinate this wider fabric.

\section{Sustainability as a 7G Design Constraint}

Sustainability is already part of the \sixthG{} design language: IMT-2030 identifies it as an overarching principle~\cite{itu_imt2030_requirements_2026}. Yet a principle is weaker than a binding constraint. For \seventhG{}, this matters because another infrastructure cycle could face less tolerance for higher absolute energy use, faster equipment turnover, or weak lifecycle evidence.

Efficiency metrics can hide growth. Bits per joule may improve while traffic, AI compute, edge-cloud deployment, site density, sensing workloads, and equipment replacement all increase. A credible \seventhG{} case should consider absolute energy consumption, embodied carbon, equipment lifetime, circularity, and grid impact, not only energy efficiency per bit. Sustainability becomes a generational gate: if \seventhG{} cannot show lower lifecycle impact than continued \sixthG{} evolution, its benefits may not justify deployment.

The AI-energy tension sharpens this point. IEA projects data-centre electricity consumption to roughly double from 485~TWh in 2025 to 950~TWh in 2030, while electricity use from AI-focused data centres is expected to triple~\cite{iea_ai_energy_2026}. AI may reduce RAN energy through optimization, but also adds telemetry, storage, training, inference, orchestration, and edge-cloud demand. The question is whether AI-native operation is net-positive across the network, compute, and energy stack.

Sustainability also extends beyond intra-network efficiency. Renewable-powered sites, battery-backed base stations, and telecom VPPs are gaining traction, but remain mostly deployment and energy-management practices rather than standardized cellular behaviours. Packetized energy management represents flexible telecom demand as schedulable energy packets shaped by renewable availability, carbon intensity, price, local constraints, and communication priorities~\cite{aijaz_2026_pem_6g}. For \seventhG{}, the question is whether standardized interfaces are needed between RAN control, edge-cloud orchestration, site energy systems, carbon signals, and grid-flexibility markets.

Lifecycle impact is important. The Global E-waste Monitor 2024 reports that e-waste increased by 82\% between 2010 and 2022 and is projected to reach 82 million tonnes by 2030~\cite{itu_ewaste_2024}. A generation that improves operational efficiency while accelerating equipment replacement may simply shift impact from electricity consumption to embodied carbon, rare materials, and waste.
A 7G sustainability framework should include at least four classes of metrics:

\begin{itemize}

    \item \textbf{Operational energy:} site power, RAN energy, cooling, transport, core, edge compute, storage, and grid-interactive flexibility.

    \item \textbf{Embodied carbon:} radios, antennas, semiconductors, batteries, masts, civil works, satellites, and data centres.

    \item \textbf{Circularity:} refurbishment, reuse, modular upgrades, repairability, lifetime extension, and recycling.

    \item \textbf{Enabled impact:} credible evidence that network capabilities reduce emissions in other sectors such as logistics, energy, transport, agriculture, and manufacturing.

\end{itemize}

Under stringent sustainability constraints, one of the most important \seventhG{} capabilities may be the ability to support new services while avoiding unnecessary hardware replacement.

%
%
%
%
%
%
%
%

\section{Can 7G survive Geopolitics?}
Global standardization has long underpinned cellular economics through device scale, spectrum harmonization, interoperable specifications, roaming, certification, and multi-vendor supply chains. \sixthG{} is already more geopolitical than earlier generations, but divergence is still mostly around the standard rather than inside the air interface: spectrum strategy, industrial policy, AI governance, supply-chain trust, security assurance, and sovereignty requirements.
This regionalization is visible in \sixthG{} policy. The United States frames \sixthG{} as a national-security, foreign-policy, and economic-prosperity priority, with emphasis on standards leadership and candidate spectrum bands~\cite{us_winning_6g_2025}. Europe links 5G-Advanced and \sixthG{} research to industrial leadership, security, privacy, sustainability, and the green and digital transitions~\cite{eu_sns_ju_2026}. China has authorized 6~GHz trial spectrum for IMT-2030 research, testing, standardization, and industrialization~\cite{china_miit_6ghz_2026}. These examples do not imply incompatible \sixthG{} standards, but show that regional policy envelopes are already diverging.

The post-\sixthG{} impact may be stronger. \sixthG{} is likely to remain anchored in 3GPP and IMT, with regional differences managed mainly through spectrum, deployment, and security profiles. A future \seventhG{} may be more exposed if it extends beyond radio into a wider fabric involving NTN, sensing, edge-cloud, AI agents, quantum-safe or quantum-enabled security, data governance, and grid-interactive energy systems. These are precisely the domains where policy is diverging fastest: AI assurance, trusted-vendor rules, sovereign cloud, data localization, security certification, supply-chain controls, spectrum sharing, and carbon reporting. Geopolitics would then shape not only deployment, but also permitted functions, accepted trust anchors, data handling, and infrastructure certification.

Three outcomes are plausible:

\begin{enumerate}
    \item \textbf{Global baseline:} one primary technical standard with regional deployment variations. This preserves scale, but assumes that spectrum, security, AI governance, and supply-chain policies remain sufficiently compatible.

    \item \textbf{Regional profiles:} common core specifications with regional differences in spectrum, security profiles, AI-control rules, vendor-trust requirements, data governance, and sustainability reporting. This preserves partial interoperability while acknowledging sovereignty constraints.

    \item \textbf{Fragmented blocs:} partially incompatible ecosystems shaped by geopolitics, industrial policy, and strategic infrastructure choices. This would weaken economies of scale and create new barriers for roaming, certification, devices, and supply chains.
\end{enumerate}


The most likely outcome is regional profiling: a common standards language with regional dialects. This would not split standards outright, but it could still shape \seventhG{} economics and design. Interoperability under policy heterogeneity should therefore be treated as a core research problem, covering policy-aware roaming, regional security profiles, explainable AI control, jurisdiction-aware data handling, supply-chain attestation, and adaptive spectrum sharing. Geopolitical viability is thus part of the \seventhG{} readiness test, not an afterthought.

\section{Toward a 7G Standardization Agenda}

The preceding sections identify candidate reasons why \seventhG{} might require standards treatment beyond \sixthG{} evolution. This section consolidates those arguments into a compact standards agenda. Table~\ref{tab:research} does not repeat the technical discussion; instead, it highlights the unresolved standardization gaps that would determine whether each candidate driver can become part of a coherent post-\sixthG{} system.
These questions show that a meaningful \seventhG{} agenda would not be confined to the air interface. The open issue is whether standards bodies can define common abstractions for autonomy, computation, trust, spectrum, sustainability, outcome assurance, and regional policy variation without turning \seventhG{} into an unbounded umbrella for every adjacent technology domain.

\begin{table}[t]
\centering
\caption{Candidate \seventhG{} standardization gaps}
\label{tab:research}
\footnotesize
\setlength{\tabcolsep}{2.5pt}
\renewcommand{\arraystretch}{1.12}
\begin{tabularx}{\columnwidth}{
>{\raggedright\arraybackslash}p{0.24\columnwidth}
>{\raggedright\arraybackslash}X
>{\raggedright\arraybackslash}X}
\toprule
\textbf{Theme} & \textbf{Standards gap} & \textbf{Possible artifact} \\
\midrule

Agentic operation &
Authority boundaries for agents outside real-time RAN control &
Intent APIs, safety envelopes, audit trails, rollback profiles \\

RF-native computing &
PHY abstraction for radios that compute as well as communicate &
Computation KPIs, calibration profiles, model broadcast hooks \\

Quantum-enabled interworking &
Service exposure beyond PQC migration &
Crypto-agility profiles, QKD/key-service interworking, trust anchors \\

Spectrum governance &
Operation across fragmented, shared, and policy-constrained bands &
Policy-aware spectrum APIs, coexistence tests, sharing databases \\

Sustainable infrastructure &
Lifecycle impact and grid interaction as system requirements &
Lifecycle KPIs, carbon-aware profiles, PEM/VPP interfaces \\

Outcome assurance &
Service models beyond slices and connectivity QoS &
Outcome KPIs, assurance telemetry, SLA and settlement primitives \\

Regionalized standards &
Interoperability under divergent policy and sovereignty constraints &
Regional profiles, policy-aware roaming, attestation frameworks \\

\bottomrule
\end{tabularx}
\vspace{-2em}
\end{table}

\section{Concluding Remarks}

Will there be a \seventhG{}? This paper has resisted both a confident yes and a dismissive no, because both answers avoid the real question: what would a seventh generation have to do that \sixthG{}, \sixthG{}-Advanced, and the surrounding wireless, cloud, and industrial ecosystems cannot? \sixthG{} already claims AI-native operation, ISAC, NTN integration, distributed cloud, sustainability, and quantum-safe security. A \seventhG{} that merely restates these ambitions with higher performance targets would not constitute a new generation in any substantive sense; it would be a new label for continued evolution.

The readiness framework developed in this paper is intended as a discipline against that outcome. Demand-led need, system-level discontinuity, coordination value, sustainability and circularity, trust, and geopolitical viability provide tests for separating generational substance from generational rhetoric. Applied to the candidate discontinuities examined here, the pattern is consistent. Agentic operation beyond real-time control, radios that compute as well as communicate, quantum-enabled interworking, policy-aware spectrum governance, grid-interactive infrastructure, outcome assurance, and regionalized standards are not automatically \seventhG{} drivers. They become credible candidates only where they require new KPIs, new interfaces, new trust models, new lifecycle constraints, or new forms of cross-domain coordination that continuous release-based evolution cannot provide.

The resulting position is deliberately conditional. \seventhG{} deserves to exist as a distinct generation only if it becomes the coordination framework for a genuinely wider post-\sixthG{} fabric spanning radio, sensing, compute, trust, energy, automation, and policy. It should instead be absorbed into \sixthG{}-Advanced, or left to adjacent ecosystems, if its proposed capabilities can be delivered without a new generational framework. The standards community does not need to answer the \seventhG{} question today. It needs the discipline to keep asking it, test by test, rather than allowing a seventh generation to arrive by default because six generations came before it.

\section*{Acknowledgment}
The theme of this paper was inspired by a panel discussion at CHEDDAR Industry Day 2026. The author thanks the moderator, Dean Bubley, and fellow panellists Paul Febvre, Syed Ali Raza Zaidi, and Kevin Holley. The views expressed in this paper are solely those of the author.
\balance

\bibliographystyle{IEEEtran}

\bibliography{cscn2026_7g_position_paper_strengthened}

\begin{thebibliography}{10}
\providecommand{\url}[1]{#1}
\csname url@samestyle\endcsname
\providecommand{\newblock}{\relax}
\providecommand{\bibinfo}[2]{#2}
\providecommand{\BIBentrySTDinterwordspacing}{\spaceskip=0pt\relax}
\providecommand{\BIBentryALTinterwordstretchfactor}{4}
\providecommand{\BIBentryALTinterwordspacing}{\spaceskip=\fontdimen2\font plus
\BIBentryALTinterwordstretchfactor\fontdimen3\font minus
  \fontdimen4\font\relax}
\providecommand{\BIBforeignlanguage}[2]{{%
\expandafter\ifx\csname l@#1\endcsname\relax
\typeout{** WARNING: IEEEtran.bst: No hyphenation pattern has been}%
\typeout{** loaded for the language `#1'. Using the pattern for}%
\typeout{** the default language instead.}%
\else
\language=\csname l@#1\endcsname
\fi
#2}}
\providecommand{\BIBdecl}{\relax}
\BIBdecl

\bibitem{itu_m2160_2023}
\BIBentryALTinterwordspacing
{ITU-R}, ``{Framework and overall objectives of the future development of IMT
  for 2030 and beyond},'' International Telecommunication Union, Recommendation
  ITU-R M.2160-0, Nov. 2023. [Online]. Available:
  \url{https://www.itu.int/rec/R-REC-M.2160}
\BIBentrySTDinterwordspacing

\bibitem{itu_imt2030_requirements_2026}
\BIBentryALTinterwordspacing
{International Telecommunication Union}, ``{IMT-2030: Technical requirements
  for the 6G future},'' ITU News, Mar. 2026, accessed: 2026-07-12. [Online].
  Available:
  \url{https://www.itu.int/hub/2026/03/imt-2030-technical-requirements-for-the-6g-future/}
\BIBentrySTDinterwordspacing

\bibitem{3gpp_rel21_2026}
\BIBentryALTinterwordspacing
{3GPP}, ``{Timeline for Release 21},'' 3GPP News, Jun. 2026, accessed:
  2026-07-12. [Online]. Available:
  \url{https://www.3gpp.org/news-events/3gpp-news/rel21-timeline}
\BIBentrySTDinterwordspacing

\bibitem{3gpp_tr38914_2026}
\BIBentryALTinterwordspacing
------, ``{First 6G RAN study approved},'' 3GPP News, Jun. 2026, {TR 38.914
  v1.0.0; accessed: 2026-07-12}. [Online]. Available:
  \url{https://www.3gpp.org/news-events/3gpp-news/6g-38914}
\BIBentrySTDinterwordspacing

\bibitem{ngmn_6g_2023}
\BIBentryALTinterwordspacing
{NGMN Alliance}, ``{6G Requirements and Design Considerations},'' Next
  Generation Mobile Networks Alliance, Tech. Rep., Feb. 2023. [Online].
  Available:
  \url{https://www.ngmn.org/publications/6g-requirements-and-design-considerations.html}
\BIBentrySTDinterwordspacing

\bibitem{kaushik_2023_isac}
\BIBentryALTinterwordspacing
A.~Kaushik \emph{et~al.}, ``{Towards Integrated Sensing and Communications for
  6G: A Standardization Perspective},'' \emph{arXiv preprint arXiv:2308.01227},
  2023. [Online]. Available: \url{https://arxiv.org/abs/2308.01227}
\BIBentrySTDinterwordspacing

\bibitem{glisic_2024_7g_enablers}
\BIBentryALTinterwordspacing
S.~Glisic, ``{Potential Enabling Technologies for 7G Networks: Survey},''
  \emph{arXiv preprint arXiv:2408.11072}, 2024. [Online]. Available:
  \url{https://arxiv.org/abs/2408.11072}
\BIBentrySTDinterwordspacing

\bibitem{lorenzo_2025_quantum_network_design_6g7g}
B.~Lorenzo, ``{Optimum 6G/7G Quantum Network Design: Survey},'' \emph{Optics
  Communications}, vol. 608, p. 131883, 2025.

\bibitem{ahmed_2026_quantum_protocol_7g}
S.~Ahmed and A.~A. Khokhar, ``{Quantum Protocol Architectures and Secure
  Control Frameworks for 7G+ Networks: Standards, Synchronization, Use Cases,
  and Challenges},'' \emph{Advanced Quantum Technologies}, 2026.

\bibitem{3gpp_ai_ml_nr}
\BIBentryALTinterwordspacing
{3GPP}, ``{Artificial intelligence and machine learning for NR air
  interface},'' 3GPP Technology Topic, 2024, accessed: 2026-07-12. [Online].
  Available: \url{https://www.3gpp.org/technologies/ai-ml-nr}
\BIBentrySTDinterwordspacing

\bibitem{etsi_isac_2025}
\BIBentryALTinterwordspacing
{ETSI}, ``{Integrated Sensing and Communications (ISAC); Use Cases and
  Deployment Scenarios},'' European Telecommunications Standards Institute,
  ETSI GR ISC 001, 2025. [Online]. Available:
  \url{https://www.etsi.org/newsroom/press-releases/2520-etsi-publishes-first-report-on-isac-use-cases-for-6g/}
\BIBentrySTDinterwordspacing

\bibitem{3gpp_ntn_2024}
\BIBentryALTinterwordspacing
{3GPP}, ``{Non-Terrestrial Networks (NTN)},'' 3GPP Technology Topic, 2024,
  accessed: 2026-07-12. [Online]. Available:
  \url{https://www.3gpp.org/technologies/ntn-overview}
\BIBentrySTDinterwordspacing

\bibitem{nist_pqc_2024}
\BIBentryALTinterwordspacing
{National Institute of Standards and Technology}, ``{NIST Releases First 3
  Finalized Post-Quantum Encryption Standards},'' NIST News, Aug. 2024,
  accessed: 2026-07-12. [Online]. Available:
  \url{https://www.nist.gov/news-events/news/2024/08/nist-releases-first-3-finalized-post-quantum-encryption-standards}
\BIBentrySTDinterwordspacing

\bibitem{gsma_zain_energy_2023}
\BIBentryALTinterwordspacing
{GSMA Foundry and Zain}, ``{Making Networks More Energy Efficient},'' GSMA
  Foundry Case Study, 2023, accessed: 2026-07-12. [Online]. Available:
  \url{https://www.gsma.com/get-involved/gsma-foundry/wp-content/uploads/2023/11/Making-Networks-More-Energy-Efficient-Zain.pdf}
\BIBentrySTDinterwordspacing

\bibitem{ericsson_energy_2022}
\BIBentryALTinterwordspacing
{Ericsson}, ``{Improving Energy Performance in 5G Networks and Beyond},''
  Ericsson Technology Review, 2022, accessed: 2026-08-11. [Online]. Available:
  \url{https://www.ericsson.com/en/reports-and-papers/ericsson-technology-review/articles/improving-energy-performance-in-5g-networks-and-beyond}
\BIBentrySTDinterwordspacing

\bibitem{itu_ewaste_2024}
\BIBentryALTinterwordspacing
{ITU and UNITAR}, ``{The Global E-waste Monitor 2024},'' International
  Telecommunication Union and United Nations Institute for Training and
  Research, 2024. [Online]. Available:
  \url{https://www.itu.int/en/ITU-D/Environment/Pages/Publications/The-Global-E-waste-Monitor-2024.aspx}
\BIBentrySTDinterwordspacing

\bibitem{wehner_2018_quantum_internet}
S.~Wehner, D.~Elkouss, and R.~Hanson, ``{Quantum Internet: A Vision for the
  Road Ahead},'' \emph{Science}, vol. 362, no. 6412, p. eaam9288, 2018.

\bibitem{yu_2026_rf_computing}
\BIBentryALTinterwordspacing
W.~Yu and V.~W.~S. Wong, ``{Analog RF Computing: A New Paradigm for
  Energy-Efficient Edge AI Over MU-MIMO Systems},'' \emph{arXiv preprint
  arXiv:2605.14331}, 2026. [Online]. Available:
  \url{https://arxiv.org/abs/2605.14331}
\BIBentrySTDinterwordspacing

\bibitem{aijaz_PQC}
A.~Kudaloor and A.~Aijaz, ``{Toward Quantum-Safe 6G: Experimental Evaluation of
  Post-Quantum Cryptography Techniques},'' \emph{IEEE Communications Standards
  Magazine}, pp. 1--8, 2026.

\bibitem{gsma_upper6ghz_2025}
\BIBentryALTinterwordspacing
{GSMA}, ``{One Year Since WRC-23: Where Are We Now with 6 GHz?}'' GSMA
  Spectrum, 2025, accessed: 2026-08-11. [Online]. Available:
  \url{https://www.gsma.com/connectivity-for-good/spectrum/one-year-since-wrc-23-where-are-we-now-with-6-ghz/}
\BIBentrySTDinterwordspacing

\bibitem{itu_wrc27_2026}
\BIBentryALTinterwordspacing
{ITU-R}, ``{Preparatory Studies for WRC-27},'' International Telecommunication
  Union, 2026, accessed: 2026-07-12. [Online]. Available:
  \url{https://www.itu.int/en/ITU-R/study-groups/rcpm/Pages/wrc-27-studies.aspx}
\BIBentrySTDinterwordspacing

\bibitem{aijaz_2026_pem_6g}
\BIBentryALTinterwordspacing
A.~Aijaz and X.~Lin, ``{Powering Net-Zero 6G: Packetized Energy Management for
  Grid-Interactive Telecom Infrastructure},'' \emph{arXiv preprint
  arXiv:2607.28111}, 2026, to appear in IEEE Conference on Standards for
  Communications and Networking (CSCN) 2026. [Online]. Available:
  \url{https://arxiv.org/abs/2607.28111}
\BIBentrySTDinterwordspacing

\bibitem{aijaz_robotics_6g_2026}
\BIBentryALTinterwordspacing
A.~Aijaz, ``{From Packets to Tasks: Rethinking 6G for Robotics},'' EngrXiv
  preprint, 2026, accessed: 2026-08-19. [Online]. Available:
  \url{https://engrxiv.org/preprint/view/7978}
\BIBentrySTDinterwordspacing

\bibitem{iea_ai_energy_2026}
\BIBentryALTinterwordspacing
{International Energy Agency}, ``{Key Questions on Energy and AI: Executive
  Summary},'' IEA Report, 2026, accessed: 2026-07-12. [Online]. Available:
  \url{https://www.iea.org/reports/key-questions-on-energy-and-ai/executive-summary}
\BIBentrySTDinterwordspacing

\bibitem{us_winning_6g_2025}
\BIBentryALTinterwordspacing
{The White House}, ``{Winning the 6G Race},'' Presidential Memorandum, Dec.
  2025, accessed: 2026-08-11. [Online]. Available:
  \url{https://www.whitehouse.gov/presidential-actions/2025/12/national-security-presidential-memorandum-nspm-8-0bda/}
\BIBentrySTDinterwordspacing

\bibitem{eu_sns_ju_2026}
\BIBentryALTinterwordspacing
{European Commission}, ``{The Smart Networks and Services Joint Undertaking},''
  Shaping Europe's Digital Future, 2026, accessed: 2026-08-11. [Online].
  Available:
  \url{https://digital-strategy.ec.europa.eu/en/policies/smart-networks-and-services-joint-undertaking}
\BIBentrySTDinterwordspacing

\bibitem{china_miit_6ghz_2026}
\BIBentryALTinterwordspacing
{Ministry of Industry and Information Technology of China}, ``{MIIT Approves 6
  GHz Trial Spectrum for 6G Development},'' State Council of the People's
  Republic of China, May 2026, accessed: 2026-08-11. [Online]. Available:
  \url{https://english.www.gov.cn/news/202605/08/content_WS69fda800c6d00ca5f9a0ad2d.html}
\BIBentrySTDinterwordspacing

\end{thebibliography}

\end{document}